# Deconvolution of VLBI Images Based on Compressive Sensing

Andriyan Bayu Suksmono
*School of Electrical Engineering and Informatics, Institut Teknologi Bandung,*
*Jl. Ganesha 10, Bandung, Indonesia*
suksmono@radar.ee.itb.ac.id, suksmono@yahoo.com

*Abstract*—Direct inversion of incomplete visibility samples in VLBI (Very Large Baseline Interferometry) radio telescopes produces images with convolutive artifacts. Since proper analysis and interpretations of astronomical radio sources require a non-distorted image, and because filling all of sampling points in the *uv*-plane is an impossible task, image deconvolution has been one of central issues in the VLBI imaging. Up to now, the most widely used deconvolution algorithms are based on least-squares-optimization and maximum entropy method. In this paper, we propose a new algorithm that is based on an emerging paradigm called compressive sensing (CS). Under the sparsity condition, CS capable to exactly reconstructs a signal or an image, using only a few number of random samples. We show that CS is well-suited with the VLBI imaging problem and demonstrate that the proposed method is capable to reconstruct a simulated image of radio galaxy from its incomplete visibility samples taken from elliptical trajectories in the *uv*-plane. The effectiveness of the proposed method is also demonstrated with an actual VLBI measured data of 3C459 asymmetric radio-galaxy observed by the VLA (Very Large Array).

I. INTRODUCTION

Aperture size defines the maximum resolution that can be achieved by a sensor, i.e., higher resolution needs larger aperture. In some cases, however, it is practically impossible to directly construct a large aperture to obtain a required resolution. This problem occurs when researchers want to achieve the resolution of radio telescope to be comparable with its optical counterparts, whose wavelength differs in more than five order of magnitudes. Construction of a single antenna with kilometers-size diameter to obtain a comparable aperture is near to impossible.

Fortunately, synthesis imaging has came to rescue. In this imaging modality, small-size antennas distributed across land are employed to obtain spatial-frequency domain samples. Additionally, Earth rotation contributes to even enlargement of the aperture. This is the basic imaging principle employed in the VLBI (Very Large Baseline Interferometry) and space-based VLBI radio telescopes.

On the other hand, it is impossible to fill out all of the discrete-spatial frequency components in the uv-plane by such a sampling mechanism. When direct inversion is performed, the result will be consisting of a true-image of the object convoluted with the PSF (Point Spread Function) of the beam, causing a severe distortion in the reconstructed image.

To reduce the distorsion, a deconvolution process should be conducted to produce a better image. Two most popular methods have been widely used by the astronomical community, i.e., the CLEAN algorithm that is based on iterative least-squares optimization [1] and the MEM (Maximum Entropy Method) [2], [3]. Variations and further developments of these two methods has been subsequently proposed by researchers, the most recent one is the multi-frequency generalizad maximum entropy method [4].

In this paper, we propose a new approach to solve the VLBI image deconvolution problem. Our method is based on an emerging paradigm called compressed sensing (or compressive sensing/sampling), hereafter will be refered to as CS [5], [6]. The Shannon sampling theorem states that a $\Delta\omega$-bandlimited signal requires at least $2\times\Delta\omega$ sampling rate for an exact reconstruction. The sampling limit in this theorem is called Nyquist rate.

Implicitly, the conventional sampling theorem assumes that the amount of information of a signal is proportional to its frequency content. In contrast, the CS proposes a new sampling paradigm in which the information content of the signal is determined by its sparsity level or by its degree of freedom. Accordingly, it is possible to construct a signal based only on a few number of samples, which is far below the Nyquist rate. At present, novel devices based on CS undergoes extensive development, among others are: compressed sensing MRI [7], single pixel camera [8], and high-speed stepped-frequency radar [9].

The rest of the paper is organized as follows. Section II explains the basic principle of synthesis imaging in VLBI. The widely used CLEAN and MEM deconvolution methods will be briefly reviewed in Section III. This Section also explains the proposed CS-based deconvolution algorithm. Section IV describes simulation of visibility measurement for a configuration of observatories, inversion to obtain a dirty image, and compressive deconvolution results of synthetic astronomical radio source. CS based reconstruction of actual VLBI data is also described in this Section. Section V concludes this paper and some remarks on CS-based synthesis imaging.

## II. A Brief on Synthesis Imaging in Radio Astronomy

In VLBI radio telescopes, an image of radio source is not captured directly by the telescope as in its optical counterpart. Instead, interferometry technique is employed to measure visibility values, which is actually two-dimensional discrete Fourier coefficients of the radio intensity distribution of an objects. Then, inverse Fourier transform is performed to obtain an image of the radio source. The following review is based on [10] and [11] with adapted notations as required in the formulation of the proposed method.

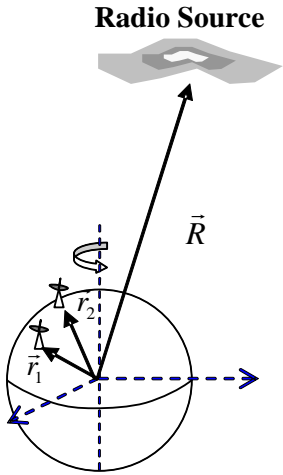

Fig. 1 A simplified imaging geometry of VLBI

Figure 1 shows a simplified geometry of the VLBI imaging. In the figure, $\vec{R}$ represents a distance vector from the observer (the antenna of radio telescope) to an object (i.e., a spatially distributed radio source), while $\vec{r}_1$ and $\vec{r}_2$ are position vectors of the corresponding observers. The source is assumed to radiate quasi-monochromatic electromagnetic wave with angular frequency ω. The electric field strength observed at position $\vec{r}$ in Earth is

$$\vec{E}(\vec{r}) = \int \frac{\vec{E}(\vec{R}) e^{j\frac{2\pi\omega|R-r|}{c}}}{|\vec{R}-\vec{r}|} dA \qquad (1)$$

where $j = \sqrt{-1}$ is the imaginary number and $c$ is the speed of light. The integration is taken over the span of the radio source $A$.

In practice, the data collected by VLBI stations is not the electric field strength, instead, the visibility $S(\vec{r}_1, \vec{r}_2)$ which corresponds to the cross-correlation between the field strengths measured by a pair of antennas that are located in $\vec{r}_1$ and $\vec{r}_2$,

$$\vec{S}(\vec{r}_1, \vec{r}_2) = \langle \vec{E}(\vec{r}_1) \cdot \vec{E}^*(\vec{r}_2) \rangle \qquad (2)$$

After substitution of (1) into (2), taking assumption of the radio-source spatial incoherence, and approximation due to $|\vec{r}| \ll |\vec{R}|$, we obtain the following formula

$$S(\vec{r}_1, \vec{r}_2) = \int s(\hat{\rho}) e^{-j\frac{2\pi\omega \vec{x}\cdot(\vec{r}_1-\vec{r}_2)}{c}} d\Omega \qquad (3)$$

where $s = \langle |\vec{E}|\rangle^2 |\vec{R}|^2$ is the intensity of the source, Ω is a solid angle spanning the object, and $\hat{\rho}$ is a unit vector in $\vec{R}$ direction, i.e., $\hat{\rho} = \vec{R}/|\vec{R}|$.

When the angular size of the radio-source is small, further approximation can be made by expanding vector $\hat{\rho}$ over the radio source $\hat{\rho} = \hat{\rho}_0 + \vec{\sigma}$, where $\hat{\rho}_0$ is orthonormal to $\vec{\sigma}$. A new coordinate system where $\hat{\rho}_0 = (0,0,1)$ can be introduced, in which the relatif position of the observer is $\vec{r}_1 - \vec{r}_2 = c(u,v,w)/\omega$ and $\hat{\rho} = (x,y,1)$. After some adjustments, then equation (3) changes into

$$S(u,v) = \iint s(x,y) e^{-j2\pi(ux+vy)} dx dy \qquad (4)$$

where $S(u,v)$ is the relative coherence function from the phase center.

Equation (4) is a Fourier relation of the true image $s(x,y)$ with the measured visibility $S(u,v)$. Accordingly, the image can be obtained by inverse Fourier transforming the visibility as follows

$$s(x,y) = \iint S(u,v) e^{j2\pi(ux+vy)} du dv \qquad (5)$$

This is the governing equation in VLBI imaging; by measuring the visibility, which is cross correlation value of electric field strength in a pair of radio-telescopes, we can obtain the image of a radio source.

Since the antennas or the VLBI stations are placed on a limited number of locations, not all of the value in the visibility function is known. Additionally, the quality of one data may differs with others, therefore weighting factors should also be incorporated into the data. The visibility in the r.h.s of Eq. (5), should be replaced by its weighted samples $B(u,v)S(u,v)$, where

$$B(u,v) = \sum_k w_k \delta(u-u_k) \delta(v-v_k) \qquad (6)$$

Due to incomplete sampling, direct inversion of (6) yields the following dirty image

$$s_D(x, y) = \sum_k w_k S(u_k, v_k) e^{j2\pi(ux+vy)} \quad (7)$$

Since a proper interpretation of the image require a non-distorted image, obtaining a good approximation of the true image from the observed data is an important task. This estimation is performed by deconvolution.

### III. VLBI IMAGE DECONVOLUTION ALGORITHMS

*A. The CLEAN and MEM algorithms*

The dirty image $s_D(x, y)$ in equation (7) is related to the true image $s(x, y)$ as

$$s_D(x, y) = s(x, y) * b(x, y) \quad (8)$$

where the dirty beam $b(x,y)$ is the (inverse) Fourier transform of $B(u,v)$ given in (6) and "*" denotes convolution. Deconvolution is a process to obtain an approximation $\hat{s}(x, y)$ of the true image $s(x,y)$ for a given dirty image $s_D(x, y)$. Since the solution of (8) is non-unique, additional prior information is required. Some reasonable assumption that has been used as priors are:

(i). Radio source has a finite support, i.e., its value is non-zero within a limited region only
(ii). Extrema in brightness, i.e., it is known that the insentity value is tnon-negative
(iii). The sky can be regarded as a collection of a small number of point sources
(iv). The sky is smooth

The most popular algorithm, CLEAN, takes prior (i), i.e., it assumes that the sky can be decomposed into a small number of sources

$$s(x, y) = \sum_{i=1}^{P} s_c \delta(x - x_i) \delta(y - y_i) \quad (9)$$

Accordingly, the dirty image $s_D(x, y)$ can be expressed as

$$s_D(x, y) = \sum_{i=1}^{P} s_C b(x - x_i, y - y_i) \quad (10)$$

Then, the parameters of the point sources is solved by least-square optimization.

In practice, this algorithm iteratively estimate the solution by finding the PSF shapes in the dirty image, which is interactively selected by the user. The CLEAN algorithm proceeds as follows:
1. Find the position and the strength of a peak in the dirty image.
2. Subtract the dirty beam multiplied by the peak strength (and a damping factor) from the dirty image.
3. Go to (1) unless the remaining peak is below a defined treshold.
4. Convolve the accumulated point source model with an idealized CLEAN beam.
5. Add the residuals of the dirty image to the CLEAN image.

Because user interaction is required, the produced cleaned image will be user dependent.

The second most popular deconvolution algorithm is the MEM. Entropy of an image $H(s)$ can be defined in various ways. A form that has a direct meaning related to information theory is

$$H(s) = -\int s(x, y) \log(s(x, y)) dx\, dy \quad (11)$$

The MEM defines the best approximation of the true image as the one that miximizes the entropy $H(s)$. In this algorithm, the maximization process is conducted through constraint optimization. At the end of the optimization process, an estimate $\hat{s}(x, y)$ of the true image is obtained.

*B. The CS-Based VLBI Deconvolution*

In the CS, reconstruction of a signal $\vec{s}$ that is sparse in a bases system $\Psi$ requires just a small number of measured samples $\vec{S}$. This subsampling process can be represented as a projection by an $M \times N$ measurement matrix $\Phi$, where $M << N$. Therefore, the observable $\hat{S}$, which is a subset of $\vec{S}$, can be expressed as follow

$$\hat{S} = \Phi \cdot \Psi \cdot \vec{s} = \Delta \cdot \vec{s} \quad (12)$$

The newly defined matrix $\Delta \equiv \Phi\Psi$ represents an over-complete basis.

Equation (12) expresses an *underdetermined* system of linear equations where the number of unknown is larger than the number the equations whose coefficients are listed in $\Delta$, therefore the solution will be non-unique. To solve this equation, CS assumes that the signal is sparse, which means that the number of the $\Psi$-domain coefficients, i.e.

$$\|\vec{s}\|_0 \equiv \sum_{n=1}^{N} |s_n|^0 \quad (13)$$

is the smallest one. Actually, minimization of (13) is a combinatorial problem that computationally intractable. When the signal is highly sparse, the solution of (13) for $L_0$ is identical to the solution of a more tractable $L_1$ problem [12], [13], by minimizing

$$\|\vec{s}\|_1 \equiv \sum_{n=1}^{N} |s_n|^1 \quad (14)$$

In fact, minimization of (15) can be recast as a convex programming problem [14], [15], whose solvers are widely available, such as the Interior Point Method.

An important issue regarding this solution is that $\Phi$ and $\Psi$ should be sufficiently incoherent. The measure of coherence between two bases $\mu(\Phi,\Psi)$ is defined as [16]:

$$\mu(\Phi,\Psi) = \max_{\phi\in\Phi, \psi\in\Psi} |\langle\phi,\psi\rangle| \qquad (15)$$

where $\phi$ and $\psi$ are column (row) vectors of $\Phi$ and $\Psi$, respectively. Recent findings in CS show that a general random basis has a high degree of incoherence with any basis, including the identity or spike bases **I**. Therefore, we can choose a random matrix as the projection bases $\Phi$. In such basis, the number of required sample $K$ is [16]

$$K \geq C \cdot \mu^2(\Phi,\Psi) \cdot F \cdot \log(N) \qquad (16)$$

where $C$ is a small constant, $F$ denotes the degree-of-freedom of the signal or the number of non-zero coefficient of the signal when represented in the sparsity bases $\Psi$.

For a suitable number of measured data $K$ given by (16), CS guarantees to recover perfectly the time domain signal through optimization

$$\min_{\vec{s}\in R^M} \|\vec{s}\|_1 \;\; s.t. \;\; \hat{S}_k = \langle\bar{\varphi}_k, \Psi\vec{s}\rangle, \; \forall k \in \{1,2,...,K\} \qquad (17)$$

where $\bar{\varphi}_k$ is a row vector of $\Phi$. In brief, the CS principle states that for a small, but sufficient, number of observations, it is possible to recover a sparse signal $\vec{s}$ from its subsamples $\hat{S}$ through $L_1$ optimization given by (17).

Regarding the image reconstruction technique from incomplete transform-domain samples [15], applying CS in VLBI deconvolution is almost straightforward. The main difference is on the measurements. Based on incomplete samples in frequency domain, i.e. the visibility $\hat{S}(u,v)$, CS estimates the true image by solving (17). The sparsity assumption of CS in VLBI is accommodated by the property (i), (iii) and (iv) that are mentioned in Sub Section III.A, i.e,
- Finite support of the source: by assuming that the spatial extension of the source is finite, the source will be sparse in spatial domain. Therefore, CS will recover the true image based on incomplete samples in (spatial) frequency domain.
- Sky as a collection of point sources: this means that the component of the image or the degree of freedom is small; therefore an incomplete sampling in frequency domain can be used to reconstruct a spatial domain image.
- The sky is smooth: smoothness of an image means that its difference is sparse. Therefore, the total variance (TV) can be used as a measure of sparsity.

In the proposed algorithm, we use the last assumption that the sky is smooth. TV minimization algorithm provided by $l_1$-*magic* can be used directly in this case. The input required by this program is the visibility values $\hat{S}(u,v)$ measured by the radio telescopes. First, a dirty image $s_D(x,y)$ is reconstructed by direct inverse Fourier transform. Then, TV minimization is applied to deconvolve the dirty image to obtain the estimate of the true image $\hat{s}(x,y)$.

IV. EXPERIMENTS AND ANALYSIS

In real life, VLBI measurement is not conducted in a totally random fashion. Locations of the sample points are determined in a well-defined manner, given the coordinate of the observers, the position of the object, and the (hour) time the measurement is performed. The purpose of the present experiments is to show the qualitative performance of CS to reconstruct (deconvolve) an image, given a set of sample points that are measured by a configuration of radio interferometers. The first experiment deals with simulated image that are sampled in discrete elliptical manners, while the second one uses actual VLBI data.

*A. Experiment with Simulated Image*

In the first experiment, a set of four hypothetical observatories and a simulated intensity distribution are used to generate the visibility data. After computing the corresponding XYZ coordinates, a set of baselines $(L_x \; L_y \; L_z)^T$ can be determined. Since baselines are difference vectors between a pair of antennas, there will be $N(N-1)/2$ number of baselines for $N$ antennas. Hence, there are 6 (six) baselines for the present case.

The location of samples in the *uv*-plane will follow a paths that are given by the following formula

$$\begin{pmatrix} u \\ v \\ w \end{pmatrix} = \frac{1}{\lambda} \begin{pmatrix} \sin H & \cos H & 0 \\ -\sin\delta\cos H & \sin\delta\sin H & \cos\delta \\ \cos\delta\cos H & -\cos\delta & \sin\delta \end{pmatrix} \begin{pmatrix} L_x \\ L_y \\ L_z \end{pmatrix} \qquad (18)$$

where $(u \; v \; w)^T$ is the *uv* (and *w*) coordinate, $\lambda = 2\pi c/\omega$ is the wavelength of the observed radio-wave, $\delta$ is the declination, and $H$ indicates time of observation. Earth rotation during observation increases $H$ and make the baselines follows elliptic trajectories in the *uv*-plane.

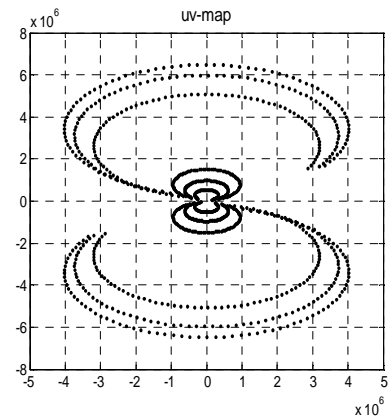

Fig. 2 UV-plane scanned by the VLBI

A simulated X-shaped radio galaxy [17], usually referred to as a model of AGN (active-galactic nuclei), is located at a particular declination and observed in a period of time. After completing the complex-conjugate values, the simulated scanning trajectories in the *uv*-plane are displayed in Fig.2. As seen on this figure, there are twelve paths of baseline indicating the location of the samples.

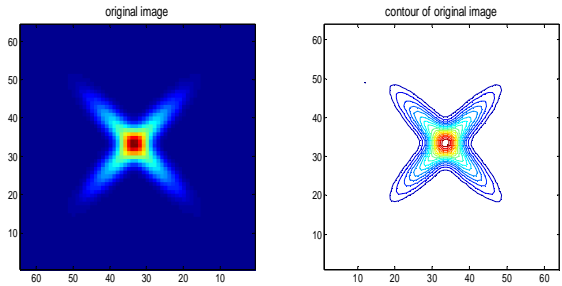

Fig. 3. Original image of simulated X-shapped radio galaxy

The original image of the simulated X-shaped radio galaxy and its corresponding contour map are displayed in Fig. 3. Based on the previously generated sample locations in the *uv*-plane, a dirty image can be reconstructed by direct Fourier inversion.

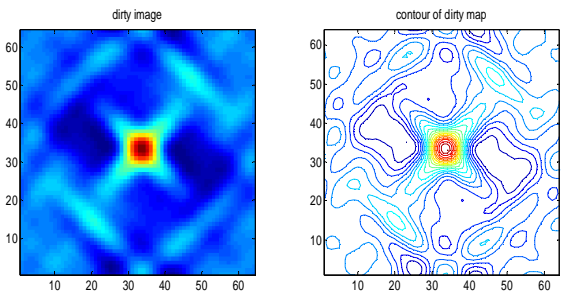

Fig. 4. Dirty image produced by direct IFFT of incomplete samples

The dirty image displayed in Fig.4 exhibits convolutionary artefacts, which are scalled-copies of X-shapped image in various places. The objective of deconvolution is to remove these artifacts to reveal the true image.

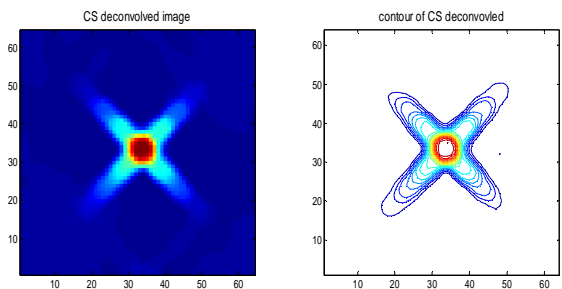

Fig. 5. Deconvolution result by CS method

At the final stage, CS reconstruction is performed based on subsamples given by the Fourier coefficients of the true image taken only in the poisitions defined in the *uv*-map. Fig.5 displays the CS-reconstruction result by $l_1$-magic, showing that the artifacts has been removed succesfully.

*B. Experiment with Real VLBI Data*

In the second experiment, we use an actual VLBI visibility data of 3C459 radio-galaxy at 4885 MHz [18], which is observed by the VLA (Very Large Array) in Socorro, New Mexico, USA. Sample points scanned by the VLA during observation are displayed in the *uv*-map shown in Fig.6. To save computation time, the data is down sampled into a grid of 128×128 size.

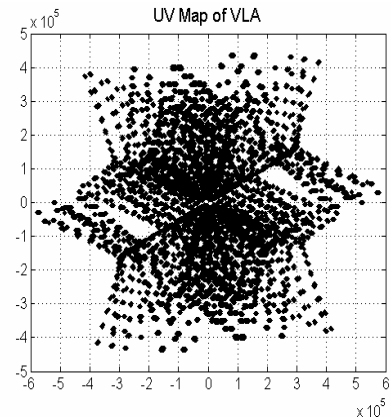

Fig. 6 The *uv*-map of VLA scanning

Figure 7 shows three-dimensional view of (a) dirty map and (b) CS reconstructed image. By comparing these figures, it is clearly observed that the dirty beam contains artefacts represented by rough surfaces around intensity's peaks. On the other hand, CS-cleaned image are more smooth and the peaks are more pronounced.

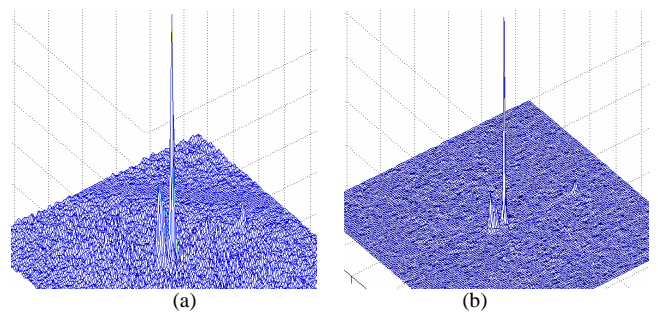

Fig. 7 Three-dimensional plot of reconstructed image:
(a) dirty image and (b) CS-cleaned

Figure 8 shows reconstructed image around an area of interest, where (a) is the dirty image and (b) CS cleaned image, while Fig.9 shows their corresponding contour plot of (a) the dirty image and (b) CS cleaned image, respectively. The contour plot shows that the artefacts in the dirty map have been removed and true image of the objects is revealed.

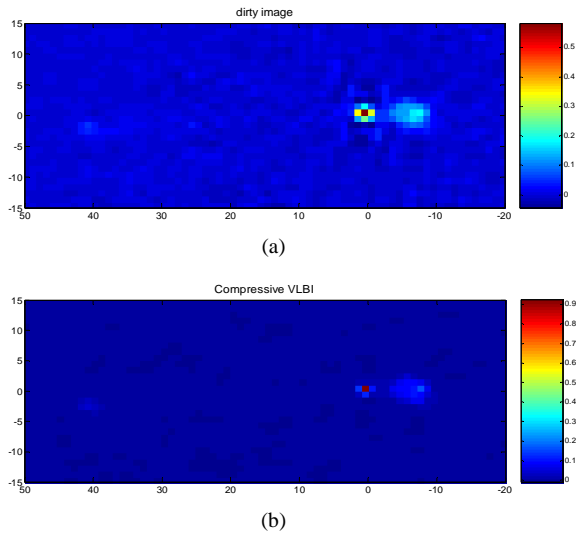

Fig. 8. Reconstructed image around an area of interrest:
(a) dirty map and (b) CS-cleaned image

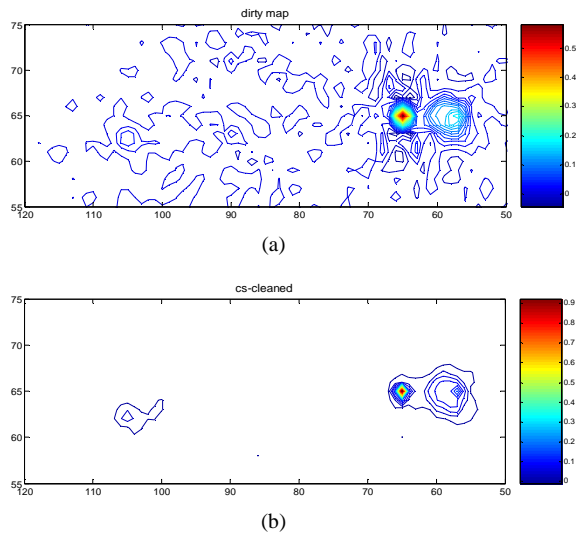

Fig. 9. Contour plot around an area of interrest at the same
contour number: (a) dirty map and (b) CS-cleaned

## V. CONCLUSIONS AND REMARKS

A new deconvolution algorithm based on compresive sensing for VLBI images has been proposed. Experiments with simulated radio-source and real VLBI data shows the potential applications of the proposed method. Although artifact removal in the dirty image is usually categorized as a deconvolution process, the proposed CS-based deconvolution is closer to faithful reconstruction process rather than a mere deconvolution. Therefore, it can be considered as a new kind of synthesis imaging that capable to perform reconstruction without deconvolution.

At the time this paper is written, a group of researchers has published related method on CS imaging techniques for radio interferometry [19], [20]. The papers use simulated image, which is random-sampled in frequency domain and show that the objective measure of the image, i.e. the SNR in this case, is increased by the CS-method. The papers also pointed out that Hogbom's CLEAN algorithm can be considered as matching pursuit (MP) algorithm, while the CS one is categorized as basis pursuit (BP). In contrast to random sampling, we use elliptical trajectories in the *uv*-plane to take the samples, which we consider to be more realistic for the VLBI case. We also validates the concept of CS reconstruction of VLBI images by an actual VLBI data. These results may complementary to each other. It seems that a new era of synthesis imaging in radio astronomy has begun.

## ACKNOWLEDGEMENTS

This work has been supported by ITB Grant of Research Division (Riset KK-ITB) 2009. We thanks ERIS (European Radio Interferometry School) to make the FITS file of 3C459 available in the internet.